\documentclass[conference]{IEEEtran}
\IEEEoverridecommandlockouts

\usepackage{cite}
\usepackage{amsmath,amssymb,amsfonts}
\usepackage{algorithmic}
\usepackage{graphicx}
\usepackage{textcomp}
\usepackage{xcolor}
\usepackage{booktabs}
\usepackage{multirow}
\usepackage{tikz}
\usetikzlibrary{shapes,arrows,positioning,calc}
\usepackage{url}
\usepackage{balance}
\usepackage{hyperref}
\hypersetup{
    hidelinks, 
}
\usepackage{orcidlink}
\usepackage{fontawesome5}
\usepackage{fancyhdr}

\tikzstyle{process} = [rectangle, minimum width=2cm, minimum height=0.7cm, text centered, draw=black, fill=blue!10]
\tikzstyle{decision} = [diamond, minimum width=2cm, minimum height=0.7cm, text centered, draw=black, fill=green!10]
\tikzstyle{io} = [trapezium, trapezium left angle=70, trapezium right angle=110, minimum width=2cm, minimum height=0.7cm, text centered, draw=black, fill=orange!10]
\tikzstyle{arrow} = [thick,->,>=stealth]
\def\BibTeX{{\rm B\kern-.05em{\sc i\kern-.025em b}\kern-.08em
    T\kern-.1667em\lower.7ex\hbox{E}\kern-.125emX}}
\begin{document}

\title{SemLink: A Semantic-Aware Automated Test Oracle for Hyperlink Verification using Siamese Sentence-BERT\\
\footnotesize
\thanks{\textsuperscript{\textsection}: Equal contribution.}
\thanks{This work was financially supported in part by the National Science and Technology Council (NSTC), Taiwan, under the following projects: NSTC 114-2221-E-002-217, NSTC 114-2622-E-A49-022, NSTC 114-2221-E-A49-210, NSTC 114-2634-F-011-002-MBK, and NSTC 114-2923-E-194-001-MY3. Additionally, this work received partial funding from the National Taiwan University (NTU) under Grant G0647, Grant 114L895501, and Grant 115L894201, within the NTU Core Consortium Project in the framework of the Higher Education Sprout Project by the Ministry of Education, Taiwan. Kuo-Hui Yeh was also financially supported in part by the Hon Hai Research Institute, Taipei, Taiwan (Project No. 114UA90042), and by the Industry-Academia Innovation School, NYCU, Taiwan (Project No. 113UC2N006).}
}
\author{
    \IEEEauthorblockN{
        Guan-Yan Yang\IEEEauthorrefmark{1}\textsuperscript{\textsection}\orcidlink{0009-0002-2539-9057},
        Wei-Ling Wen\IEEEauthorrefmark{1}\textsuperscript{\textsection}\orcidlink{0009-0003-2511-4389},
        Shu-Yuan Ku\IEEEauthorrefmark{1}\textsuperscript{\textsection},
        Farn Wang\IEEEauthorrefmark{1}\orcidlink{0000-0002-0425-6500},
        Kuo-Hui Yeh\IEEEauthorrefmark{2}\IEEEauthorrefmark{3}\orcidlink{0000-0003-0598-761X}
    }
    \IEEEauthorblockA{\IEEEauthorrefmark{1}National Taiwan University, Taipei, Taiwan}
    \IEEEauthorblockA{\IEEEauthorrefmark{2}National Yang Ming Chiao Tung University, Hsinchu, Taiwan}
    \IEEEauthorblockA{\IEEEauthorrefmark{3}National Dong Hwa University, Hualien, Taiwan}
    
    \IEEEauthorblockA{
       \{f11921091, r12921046, farn\}@ntu.edu.tw, w26662776@gmail.com,  khyeh@nycu.edu.tw
    }
}
    


\maketitle

\thispagestyle{fancy}
\fancyhf{}
  \renewcommand{\headrulewidth}{0pt} 
  \renewcommand{\footrulewidth}{0pt} 
  
  \fancyhead[L]{%
    \parbox{\textwidth}{\centering 
      \vspace*{-0.1in} 
      \fontsize{6}{9}\selectfont
      \textsf{ 
      Preprint. This article has been accepted at the 19th IEEE International Conference on Software Testing, Verification and Validation (ICST) 2026. This is the author's version which has not been fully edited and content may change prior to final publication.
      }
    }
  }

\renewcommand{\headrulewidth}{0pt}

\begin{abstract}
Web applications rely heavily on hyperlinks to connect disparate information resources. However, the dynamic nature of the web leads to ``link rot,'' where targets become unavailable, and more insidiously, ``semantic drift,'' where a valid HTTP 200 connection exists, but the target content no longer aligns with the source context. Traditional verification tools, which primarily function as crash oracles by checking HTTP status codes, often fail to detect semantic inconsistencies, thereby compromising web integrity and user experience. While Large Language Models (LLMs) offer semantic understanding, they suffer from high latency, privacy concerns, and prohibitive costs for large-scale regression testing. 
In this paper, we propose SemLink, a novel automated test oracle for semantic hyperlink verification. SemLink leverages a Siamese Neural Network architecture powered by a pre-trained Sentence-BERT (SBERT) backbone to compute the semantic coherence between a hyperlink's source context (anchor text, surrounding DOM elements, and visual features) and its target page content. To train and evaluate our model, we introduce the Hyperlink-Webpage Positive Pairs (HWPPs) dataset, a rigorously constructed corpus of over 60,000 semantic pairs. Our evaluation demonstrates that SemLink achieves a Recall of 96.00\%, comparable to state-of-the-art LLMs (GPT-5.2), while operating approximately 47.5 times faster and requiring significantly fewer computational resources. This work bridges the gap between traditional syntactic checkers and expensive generative AI, offering a robust and efficient solution for automated web quality assurance.
\end{abstract}


\section{Introduction}
\label{sec:introduction}

Hyperlinks are the fundamental structural element of the World Wide Web, serving as the primary mechanism for information retrieval and navigation. For modern web applications, maintaining the integrity of these links is a critical component of software quality assurance~\cite{BALSAM2025112186}. However, the decentralized and mutable nature of the web introduces a pervasive degradation phenomenon known as ``link rot'' ~\cite{chapekis2024online}. 

Traditionally, link rot refers to hyperlinks that point to resources that have permanently disappeared, typically returning 404 Not Found or 410 Gone HTTP status codes. Detecting these failures is a trivial task for automated crawlers and is a standard practice in Continuous Integration/Continuous Deployment (CI/CD) pipelines. However, a more subtle and challenging failure mode exists: \textit{semantic drift} (or ``soft link rot''). In this scenario, a hyperlink returns a successful HTTP 200 OK status, but the content of the target webpage has changed significantly or is entirely irrelevant to the context in which the link was originally embedded. Examples include domain parking pages, soft 404s (custom error pages returning status 200), or generic homepages that no longer host the specific referenced content.

\subsection{The Test Oracle Problem}
In the context of software testing, this presents a classic \textit{Test Oracle Problem}. A test oracle is a mechanism for determining whether the output of a system (the target page content) matches the expected behavior~\cite{barr2015oracle} (the semantic promise made by the anchor text). Traditional tools like W3C Link Checker ~\cite{deadlinkchecker} or Screaming Frog ~\cite{screamingfrog} function as \textit{crash oracles}—they can only detect if the retrieval process crashes. They lack the semantic understanding required to act as \textit{functional oracles} that verify content relevance.

While Large Language Models (LLMs) offer semantic capabilities ~\cite{liu2025relevance, Pradel2026Testora}, integrating them into high-frequency regression testing introduces critical bottlenecks: 1) \textbf{Cost:} Enterprise-scale verification incurs prohibitive API or infrastructure expenses; 2) \textbf{Latency:} Generative inference is orders of magnitude slower than discriminative models; and 3) \textbf{Determinism:} Stochastic outputs lead to flaky tests. 
Furthermore, deploying LLMs introduces unique security risks related to instruction tuning. Specifically, LLMs are susceptible to prompt injection ~\cite{wei2024jailbroken, yang2025artperception}, where manipulated input data hijacks the model's control flow, potentially causing context leakage or unauthorized behavior. In contrast, SemLink utilizes a fixed discriminative architecture that processes inputs strictly as feature vectors. 

\subsection{Proposed Approach: SemLink}
To address these challenges, we present \textbf{SemLink}, a specialized deep learning framework designed to serve as an automated semantic oracle for hyperlink verification. Unlike general-purpose LLMs, SemLink utilizes a Siamese Neural Network (SNN) architecture optimized for semantic similarity ranking. We employ a pre-trained Sentence-BERT (SBERT) ~\cite{reimers2019sentence} model to generate dense vector representations of both the hyperlink context (source) and the webpage content (target). These embeddings are processed through a weight-sharing neural network trained via Binary Cross-Entropy (BCE) Loss to optimize the alignment between the predicted probabilities and the ground-truth labels.

Crucially, our approach goes beyond simple anchor-text matching. We implement a heuristic-weighted feature extraction pipeline that captures the ``Side-Text'' (surrounding DOM elements) and ``Image Text'' (via OCR and attribute extraction), acknowledging that modern web design often conveys semantic intent through context and visual elements rather than anchor text alone.

\subsection{Contributions}
This paper makes the following contributions to the field of automated web testing:
\begin{itemize}
    \item \textbf{A Semantic Test Oracle:} We propose a Siamese SBERT-based architecture that effectively automates the verification of semantic consistency between hyperlinks and target pages, detecting failures that traditional HTTP checkers miss.
    \item \textbf{The HWPPs Dataset:} We construct and release the Hyperlink-Webpage Positive Pairs dataset, containing over 60,000 curated pairs from 500 diverse real-world websites. This dataset fills a gap in available resources for training semantic relevance models in the software engineering domain.
    \item \textbf{Heuristic Context Extraction:} We introduce a DOM-based weighting strategy for extracting hyperlink context, demonstrating that surrounding text and visual attributes significantly improve verification accuracy compared to anchor text alone.
    \item \textbf{Efficiency Analysis:} We provide a rigorous empirical evaluation showing that SemLink achieves 96.00\% Recall—within 2.3\% of GPT-5.2—while operating 47.5 times faster, making it a viable candidate for integration into real-time CI/CD testing environments.
\end{itemize}

The source code will be made available upon the publication of the extended journal version of this work.

\section{Related Work}
\label{sec:related}

The challenge of hyperlink verification resides at the intersection of web engineering, information retrieval, and automated software testing. In this section, we review the evolution of link analysis from structural connectivity to semantic consistency, situate our approach within the emerging field of neural test oracles, and discuss the computational trade-offs necessitating our specific architectural choices.

\subsection{From Connectivity Checks to Semantic Consistency}
Hyperlink analysis has historically focused on topological structure and binary connectivity. Foundational studies by Park and Thelwall ~\cite{park2003hyperlink} and Henzinger ~\cite{henzinger2001hyperlink} established frameworks for analyzing web graph connectivity, treating links primarily as navigational edges. Consequently, the industry standard for handling ``link rot''---the permanent unavailability of resources---has relied on \textit{Crash Oracles}. Tools such as the W3C Link Checker and Screaming Frog ~\cite{deadlinkchecker, screamingfrog} detect ``hard'' failures (e.g., HTTP 404/410) but remain blind to ``soft'' failures where a server returns a success code (HTTP 200) despite content degradation or irrelevance---a phenomenon known as \textit{semantic drift} ~\cite{chapekis2024online}.

To address broken links, research has largely pivoted towards \textit{repair} rather than \textit{validation}. Martinez-Romo and Araujo ~\cite{martinez2012updating, martinez2008recommendation} introduced information retrieval techniques to suggest replacement URLs for broken links. More recently, Qi et al. proposed SEMTER ~\cite{qi2023semantic}, a deep learning-based repair technique that leverages UI context to fix broken interactions, while Wen et al. ~\cite{Wen24Enhancing} further enhanced script repair by integrating UI structural and visual information. Similarly, Mahajan et al. ~\cite{Mahajan17Automated} applied search-based techniques to automatically repair cross-browser layout issues. While these approaches effectively mitigate the aftermath of a crash, they act retroactively. They do not solve the \textit{Test Oracle Problem} ~\cite{barr2015oracle} for active links that are technically functional but semantically invalid.

Early attempts to incorporate semantic validation, such as those by Blustein et al. ~\cite{blustein1997methods} and Lin et al. ~\cite{lin2017using}, utilized statistical measures like TF-IDF and Latent Semantic Indexing. However, these statistical methods often struggle to capture the contextual nuances of short, ambiguous anchor texts (e.g., ``Read More''), highlighting the need for deeper semantic understanding.

\subsection{Neural Test Oracles for Web Applications}
The limitation of manual assertions has driven the development of \textit{Neural Test Oracles}, which learn expected system behavior from data. In the domain of GUI testing, deep learning has successfully bridged the semantic gap between code and visual perception. For instance, LabelDroid by Chen et al. ~\cite{chen2020unblind} utilizes deep learning to predict natural language labels for mobile UI components, enabling semantic interactions. Similarly, Nass et al. ~\cite{nass2023similarity} and Kirinuki et al. ~\cite{Kirinuki2021NLP} have applied NLP and similarity metrics to robustly identify and localize web elements despite UI evolution. This trend reflects a broader shift in software testing towards evaluating linguistic and semantic capabilities, as explored in recent works by Huang et al. ~\cite{Huang22AEON} and Lee et al. ~\cite{Lee24Automated}.

However, a critical distinction exists between \textit{Element Identification} and \textit{Hyperlink Verification}. The aforementioned works focus on resolving a single element within a single page context. In contrast, hyperlink verification is inherently relational: it requires modeling the semantic coherence between two distinct resources---the source context ($H_{info}$) and the target content ($P_{info}$). Existing UI oracles do not explicitly model this cross-document semantic flow, leaving a gap for a specialized oracle designed to verify the ``promise'' made by a navigational anchor.

\subsection{Efficient Semantic Similarity in Testing Pipelines}
To model the relationship between source and target text, recent advancements in Transformer models, specifically BERT ~\cite{devlin2018bert}, offer powerful contextual embeddings. However, integrating these Large Language Models (LLMs) into high-frequency regression testing introduces significant efficiency bottlenecks.

As noted in the survey by Han et al.~\cite{han2021survey}, while deep learning excels at short text similarity, the standard ``Cross-Encoder'' architecture of BERT is computationally prohibitive for pair-wise tasks. It requires feeding every potential pair into the network simultaneously to compute full self-attention, making it unscalable for validating thousands of links in a CI/CD pipeline ~\cite{reimers2019sentence}. Generative LLMs (e.g., GPT-5.2) further exacerbate this issue with high latency and non-deterministic outputs.

To resolve the tension between semantic accuracy and operational efficiency, Reimers and Gurevych introduced Sentence-BERT (SBERT) ~\cite{reimers2019sentence}. By employing a Siamese architecture with shared weights (Bi-Encoders), SBERT decouples the processing of inputs, allowing for the pre-computation of embeddings and rapid cosine similarity comparisons. \textit{SemLink} builds upon this architectural efficiency. We adapt the Siamese paradigm specifically for web navigation, effectively bridging the gap between fast but shallow syntactic checkers and powerful but slow generative models.

\section{Preliminaries}
\label{sec:preliminaries}

To facilitate the detailed description of our methodology in subsequent sections, we first define the core technologies and mathematical frameworks utilized in SemLink: the Document Object Model (DOM) for feature extraction, and Siamese Neural Networks with BCE Loss for semantic evaluation.

\subsection{The Document Object Model (DOM)}
The DOM represents HTML documents as a tree structure, providing essential topological context for hyperlink verification ~\cite{whatwg_dom}. Key components include: \textit{Element Nodes} representing HTML tags (e.g., \texttt{<a>}, \texttt{<div>}) targeted by our extraction algorithm; \textit{Attribute Nodes} providing metadata such as \texttt{href} and \texttt{alt}, crucial for identifying link targets and context; and \textit{Text Nodes} containing the visible content.

This hierarchical structure underpins our weighting strategy (Section \ref{sec:methodology}). By analyzing sibling and parent nodes, we extract ``surrounding context'' (Side-Text) to disambiguate generic anchor texts.

\subsection{Siamese Neural Networks}
Siamese Neural Networks (SNN) are a class of neural architectures that contain two or more identical sub-networks ~\cite{chicco2021siamese}. These sub-networks have the same configuration with the same parameters and weights. Parameter updating is mirrored across both sub-networks.

SNNs are specifically designed for comparison tasks. Unlike classifying an input into discrete categories(e.g., ``Dog'' vs. ``Cat''), an SNN accepts two distinct inputs and outputs a similarity score representing how close they are in a learned feature space. Our system utilizes this architecture to map semantically similar hyperlink-webpage pairs to close points in the embedding space, and dissimilar pairs to distant points.





\section{The HWPPs Dataset Construction}
\label{sec:dataset}

A significant barrier to advancing semantic hyperlink verification is the lack of public benchmarks. While datasets for broken links exist (e.g., lists of 404 URLs), there is no standard corpus for \textit{semantic consistency}—pairs of (hyperlink, target page) that are known to be semantically relevant. To train our Siamese Neural Network, we constructed the \textbf{Hyperlink-Webpage Positive Pairs (HWPPs)} dataset. This section details our collection methodology, seed selection strategy, and statistical properties.

\subsection{Data Collection Methodology}
We operate under the \textit{Maintenance Assumption}: Hyperlinks found on high-traffic, actively maintained websites (e.g., major news portals, government sites) are, with high probability, semantically correct (Positive Pairs). By crawling these sites, we can generate a large-scale dataset of valid $(H_{info}, P_{info})$ pairs to serve as ground truth for "Relevant" classes.

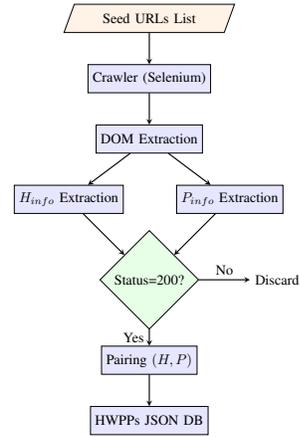
\begin{figure}[t]
\centering
\resizebox{0.45\columnwidth}{!}{
    \begin{tikzpicture}[node distance=1.5cm, auto]
        \node (input) [io] {Seed URLs List};
        \node (crawl) [process, below of=input] {Crawler (Selenium)};
        \node (extract) [process, below of=crawl] {DOM Extraction};
        
        \node (h_extract) [process, below of=extract, xshift=-2cm] {$H_{info}$ Extraction};
        \node (p_extract) [process, below of=extract, xshift=2cm] {$P_{info}$ Extraction};
        
        \node (filter) [decision, below of=extract, yshift=-2cm] {Status=200?};
        \node (pair) [process, below of=filter, yshift=-0.5cm] {Pairing $(H, P)$};
        \node (db) [process, below of=pair] {HWPPs JSON DB};
    
        \draw [arrow] (input) -- (crawl);
        \draw [arrow] (crawl) -- (extract);
        \draw [arrow] (extract) -- (h_extract);
        \draw [arrow] (extract) -- (p_extract);
        \draw [arrow] (h_extract) -- (filter);
        \draw [arrow] (p_extract) -- (filter);
        \draw [arrow] (filter) -- node[anchor=east] {Yes} (pair);
        \draw [arrow] (filter) -- node[anchor=south] {No} +(2.5,0) node[anchor=west] {Discard};
        \draw [arrow] (pair) -- (db);
    \end{tikzpicture}
}
\caption{The automated data collection pipeline for the HWPPs dataset. The system crawls seed URLs, extracts hyperlink context ($H_{info}$) and target page content ($P_{info}$), filters for accessibility, and stores valid pairs.}
\label{fig:collection_pipeline}
\end{figure}

The data collection pipeline is illustrated in Fig. \ref{fig:collection_pipeline}. The process consists of four stages.

\subsubsection{Seed Selection}
To ensure domain diversity and reduce bias, we selected 381
seed URLs from five distinct categories: News Media (e.g., BBC, CNN), E-Commerce (e.g., Amazon, eBay), Educational Institutions (e.g., University homepages), Government Portals, and Tech Blogs. The selection criteria required that the website be updated daily or weekly, ensuring the links were likely to be "fresh" and valid. The linguistic distribution was designed to be bilingual, comprising approximately 75\% Chinese and 25\% English sources.

\subsubsection{Crawling and Extraction}
We developed a specialized crawler using Python's \texttt{Selenium} framework to handle dynamic JavaScript content. For every seed URL, the crawler:
\begin{enumerate}
    \item Identifying all anchor tags (\texttt{<a>}) in the DOM.
    \item Filtering out non-navigational links (e.g., \texttt{javascript:void(0)}, \texttt{mailto:}, anchor jumps \texttt{\#section}).
    \item Extracting the Source Context ($H_{info}$) including anchor text, image attributes, and surrounding DOM text.
    \item Visiting the \texttt{href} target URL. If the response status is HTTP 200, extracting the Target Content ($P_{info}$) including page title, headers (\texttt{<h1>}-\texttt{<h3>}), and main body keywords.
\end{enumerate}

\subsubsection{Pairing and Cleaning}
Each valid hyperlink $H$ and its successfully retrieved target page $P$ form a positive pair $(H, P)$. We applied strict cleaning rules: pairs with empty anchor text (and no image alt text) or empty target page bodies were discarded. This resulted in a final dataset of \textbf{63,870 pairs} collected from 381 unique source domains.

\subsection{Dataset Statistics and Structure}
Our HWPPs dataset is stored in a structured JSON format. This dataset is split into training (85\%, $\approx$55,000 pairs) and validation (15\%, $\approx$8,870 pairs) sets. Each entry contains rich metadata:
\begin{itemize}
    \item \textbf{Link Info:} URL, Anchor Text, Link Type (Text/Image), Side-Text (List of 5 nearest neighbors), Image OCR text (if applicable).
    \item \textbf{Webpage Info:} URL, HTTP Status, Page Title, Header List ($H1...H3$), Extracted Keywords (via TextRank).
\end{itemize}

\subsection{Semantic Validity Analysis}
To verify our "Positive Pair" assumption, we conducted a preliminary analysis using a standard pre-trained Sentence-BERT model without fine-tuning. We computed the cosine similarity between the anchor text and the target page title for all pairs.
The analysis revealed that 85\% of the collected pairs had a base similarity score $>0.5$, and nearly 50\% had a score $>0.9$. This strong baseline semantic correlation confirms that our collection strategy successfully captured relevant pairs, making HWPPs a robust foundation for training a more discriminative Siamese network. The remaining 15\% with lower scores often represented "generic" links (e.g., "Read More") where simple cosine similarity fails, highlighting the necessity for our proposed context-aware training approach.

\section{Methodology: Feature Extraction}
\label{sec:methodology}

The core hypothesis of SemLink is that a hyperlink's semantic intent is rarely contained solely within its anchor text. "Read More" buttons, icon-only links, and navigational menus rely heavily on surrounding context and visual cues to convey meaning. Therefore, our methodology begins with a robust feature extraction pipeline that transforms raw HTML into rich semantic representations.

\begin{figure}[tp]
\centering
\resizebox{0.55\columnwidth}{!}{
    \begin{tikzpicture}[node distance=1.2cm, auto]
        \tikzstyle{block} = [rectangle, draw, fill=blue!10, text width=6em, text centered, rounded corners, minimum height=3em]
        \tikzstyle{decision} = [diamond, draw, fill=green!10, text width=4em, text badly centered, node distance=2cm, inner sep=0pt]
        \tikzstyle{line} = [draw, -latex']
        \tikzstyle{cloud} = [draw, ellipse, fill=red!10, node distance=2.5cm, minimum height=2em]
    
        \node [cloud] (start) {HTML Source};
        \node [block, below of=start] (find_a) {Find $\langle a \rangle$ Tag};
        \node [decision, below of=find_a] (is_img) {Has Image?};
        
        \node [block, left of=is_img, node distance=3cm] (anchor) {Extract Anchor Text};
        
        \node [block, right of=is_img, node distance=3cm] (ocr) {OCR \& Alt Attr.};
        
        \node [block, below of=is_img, node distance=2cm] (sidetext) {DOM Traversal (Side-Text)};
        
        \node [block, below of=sidetext, node distance=1.5cm] (target) {Visit Target URL};
        \node [block, below of=target, node distance=1.5cm] (content) {Extract Title, h1-h3, Keywords};
        
        \path [line] (start) -- (find_a);
        \path [line] (find_a) -- (is_img);
        \path [line] (is_img) -- node [near start] {No} (anchor);
        \path [line] (is_img) -- node [near start] {Yes} (ocr);
        \path [line] (anchor) |- (sidetext);
        \path [line] (ocr) |- (sidetext);
        \path [line] (sidetext) -- (target);
        \path [line] (target) -- (content);
    \end{tikzpicture}
}
\caption{The SemLink Feature Extraction Pipeline. The system handles both text-based and image-based links, augmenting them with contextual ``Side-Text'' derived from the DOM tree structure.}
\label{fig:extraction_logic}
\end{figure}
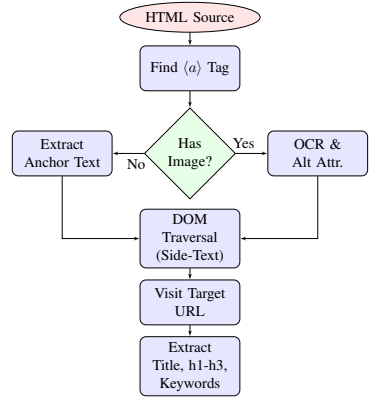

The extraction process is divided into two parallel streams: Source Hyperlink Context Extraction ($H_{info}$) and Target Webpage Content Extraction ($P_{info}$). The workflow is visualized in Fig. \ref{fig:extraction_logic}.

\subsection{Source Context Extraction ($H_{info}$)}
For a given hyperlink $l$, we define its semantic representation $H_{info}$ as a tuple of features:
\begin{equation}
H_{info}(l) = \{T_{anchor}, T_{img}, S_{context}\}
\end{equation}

\subsubsection{Anchor Text ($T_{anchor}$)}
The primary signal is the text enclosed within the \texttt{<a>} tags. While sufficient for descriptive links (e.g., ``Contact Support''), it is often insufficient for generic navigational elements.

\subsubsection{Visual-Semantic Extraction ($T_{img}$)}
Modern web design frequently utilizes clickable images or icons without explicit text. If the anchor tag encloses an \texttt{<img>} element, we extract information from three sources to construct $T_{img}$:
\begin{itemize}
    \item \textbf{Attributes:} The \texttt{alt} and \texttt{title} attributes often contain developer-provided descriptions.
    \item \textbf{Optical Character Recognition (OCR):} We employ the EasyOCR library to extract embedded text from image files. This is crucial for promotional banners where the semantic payload (e.g., "Summer Sale 50\% Off") is pixel-based rather than text-based.
    \item \textbf{Icon Classification:} For icon-only links (e.g., a magnifying glass for search), we utilize pre-trained classifications to map visual symbols to semantic concepts, though experimental ablation showed OCR and attributes to be the dominant factors.
\end{itemize}

\subsubsection{Heuristic Side-Text Extraction ($S_{context}$)}
This is the novel contribution of our extraction logic. To resolve ambiguity in generic links, we implement a DOM Traversal Heuristic. We posit that the semantic meaning of a link is partially inherited from its neighbors.
The algorithm traverses:
\begin{enumerate}
    \item \textbf{Siblings:} Text nodes immediately preceding or following the anchor tag within the same parent container.
    \item \textbf{Parent:} Text nodes belonging to the parent container, often providing the category for a list of links.
\end{enumerate}
We extract up to $k=5$ distinct text snippets surrounding the anchor. To reflect the diminishing relevance of distant text, we assign position-dependent importance weights during the similarity evaluation phase (detailed in Section \ref{sec:model_architecture}). For example, in a news card layout, the headline (sibling) gives meaning to the ``Read More'' button (anchor).

\subsection{Target Content Extraction ($P_{info}$)}
Once the source context is established, SemLink verifies the target. Upon receiving an HTTP 200 response, we parse the target DOM to construct $P_{info}$:
\begin{equation}
P_{info}(p) = \{T_{title}, T_{headers}, K_{body}\}
\end{equation}

\begin{itemize}
    \item \textbf{Page Title ($T_{title}$):} The \texttt{<title>} tag provides the highest-level semantic summary of the page.
    \item \textbf{Structural Headers ($T_{headers}$):} We extract all \texttt{<h1>}, \texttt{<h2>}, and \texttt{<h3>} tags. These headers typically outline the document's main topics and are less noisy than body text.
    \item \textbf{Content Keywords ($K_{body}$):} Extracting the entire body text often introduces excessive noise (navigation menus, footers). Instead, we extract the main content block and apply the \textbf{TextRank} algorithm~\cite{mihalcea2004textrank} to identify the top-$N$ keywords. This provides a dense semantic summary of the target page without exceeding the token limits of downstream embedding models.
\end{itemize}

This rigorous extraction process ensures that the subsequent neural network receives high-quality, semantically dense inputs, distinguishing SemLink from tools that look only at URL strings or status codes.

\section{Methodology: The SemLink Model Architecture}
\label{sec:model_architecture}

While feature extraction prepares the raw data, the core intelligence of SemLink resides in its neural architecture. We employ a Siamese Neural Network (SNN) designed to learn a similarity function between the source context $H_{info}$ and the target content $P_{info}$. Unlike standard classification models that output a class label, our SNN outputs a continuous similarity score, allowing for a tunable threshold that can trade off precision and recall based on testing requirements.

\begin{figure}[tp]
\centering
\resizebox{0.65\columnwidth}{!}{ 
    \begin{tikzpicture}[node distance=1.4cm, auto]
        \tikzstyle{layer} = [rectangle, draw, fill=blue!10, text width=8em, text centered, minimum height=2em]
        \tikzstyle{op} = [rectangle, rounded corners=16pt, draw, fill=orange!20, text centered, inner sep=5pt]
        \tikzstyle{input} = [trapezium, trapezium left angle=70, trapezium right angle=110, draw, fill=green!10, text width=4.5em, text centered]
    
        \node [input] (inputA) {Source Text ($T_H$)};
        \node [layer, below of=inputA] (sbertA) {SBERT Backbone};
        \node [layer, below of=sbertA] (embedA) {Embedding Layer};
    
        \node [input, right of=inputA, node distance=5.8cm] (inputB) {Target Text ($T_P$)};
        \node [layer, below of=inputB] (sbertB) {SBERT Backbone};
        \node [layer, below of=sbertB] (embedB) {Embedding Layer};
    
        \draw[<->, dashed] (sbertA) -- node {Shared Weights} (sbertB);
        \draw[<->, dashed] (embedA) -- node {Shared Weights} (embedB);
    
        \node [op, below of=embedA, xshift=2.9cm, align=center] (diff) {Semantic Difference \\ Vector ( $d = |v_H-v_P|$ )};
        \node [layer, below of=diff] (MLP) {Multi-Layer \\Perceptron(MLP)};
        \node [below of=MLP] (output) {Similarity Score $\hat{y}$};
    
        \draw [arrow] (inputA) -- (sbertA);
        \draw [arrow] (sbertA) -- (embedA);
        \draw [arrow] (embedA) |- (diff);
    
        \draw [arrow] (inputB) -- (sbertB);
        \draw [arrow] (sbertB) -- (embedB);
        \draw [arrow] (embedB) |- (diff);
    
        \draw [arrow] (diff) -- (MLP);
        \draw [arrow] (MLP) -- (output);
    \end{tikzpicture}
}
\caption{The SemLink Siamese Network Architecture. Two inputs are processed by identical SBERT backbones and embedding layer with shared weights. Then, semantic difference vector is passed through MLP to predict semantic similarity.}
\label{fig:snn_architecture}
\end{figure}
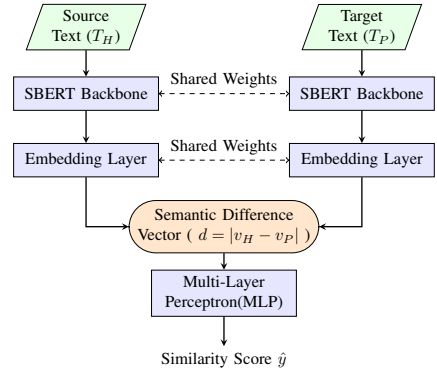

The architecture is visualized in Fig. \ref{fig:snn_architecture}. It consists of three primary stages: the Sentence-BERT (SBERT) Backbone, the Siamese Comparator, and the Aggregation Logic.

\subsection{Siamese SBERT Backbone}
The backbone of our system is a pre-trained Sentence-BERT model ~\cite{reimers2019sentence}. We utilize the \texttt{distiluse-base-multilingual-cased-v2} variant\footnote{Source from: \url{https://huggingface.co/sentence-transformers/distiluse-base-multilingual-cased-v2}.} for its balance of speed and accuracy. Let $T_H$ be a textual component from the hyperlink (e.g., anchor text) and $T_P$ be a textual component from the target page (e.g., page title). 

The feature extraction process proceeds in two stages. First, the SBERT backbone encodes these variable-length text strings into 512-dimensional intermediate vectors, denoted as $\mathbf{e}_h, \mathbf{e}_p$. Subsequently, a linear embedding layer projects these vectors into a lower-dimensional feature space, resulting in the final 128-dimensional representations $\mathbf{v}_H, \mathbf{v}_P$. Formally, we define the encoding function $M(\cdot)$ to represent this composite transformation. The final embeddings are thus obtained as:
\begin{equation}
\mathbf{v}_H = M(T_H), \mathbf{v}_P = M(T_P)
\end{equation}

Crucially, the weights of $M$ are shared across both inputs, ensuring that the embedding space is consistent for both source and target text. This "twin" structure is fundamental to Siamese networks, enforcing the constraint that similar semantic concepts must map to similar geometric locations regardless of whether they appear in a link or a webpage.

\subsection{Comparator and Classification Head}
While cosine similarity is the standard metric for SBERT embeddings, we found that fine-tuning a similarity estimator yields superior performance for the specific domain of web navigation.
To capture the semantic relationship, we first compute the element-wise absolute difference between the embedding vectors  $\mathbf{v}_H$ and $\mathbf{v}_P$: \begin{equation}
d = |\mathbf{v}_H - \mathbf{v}_P|
\end{equation}

The difference vector $d$ serves as the input to a Multi-Layer Perceptron (MLP) consisting of three fully connected layers. The first two layers are configured with 128 hidden units each, followed by Dropout regularization and ReLU activation to capture non-linear dependencies. The final layer projects these 128-dimensional features to a scalar value, which is then passed through a Sigmoid activation function. This explicitly constrains the final similarity output $\hat{y}$ to the interval $[0, 1]$, representing the predicted probability of semantic relevance.

\subsection{Optimization Objective}
\normalsize
To effectively train our model, we formulated a generalized loss framework that considers both geometric constraints and probabilistic calibration. Specifically, we explored the combination of Triplet Margin Loss ~\cite{schroff2015facenet} and Binary Cross-Entropy (BCE) Loss ~\cite{bishop2006pattern}. 

\normalsize
While Triplet loss aims to optimize relative distances in the embedding space, BCE Loss focuses on the absolute output probability. For a given anchor input $A$ (hyperlink), a positive sample $P$ (correct webpage), and a negative sample $N$ (irrelevant webpage), the total loss function $\mathcal{L}_{total}$ is defined as:
\begin{equation}
\mathcal{L}_{total} = \lambda_1 \mathcal{L}_{tri}(A, P, N) + \lambda_2 [\mathcal{L}_{BCE}(A, P) + \mathcal{L}_{BCE}(A, N)]
\label{total_loss}
\end{equation}
where $\lambda_1$ and $\lambda_2$ are hyperparameters weighting the contribution of each component. Let $\mathbf{v}_x$ denote the embedding vector for input $x$. The Triplet Margin Loss enforces a separation margin $\alpha$ between the positive and negative pairs:
\begin{equation}
\mathcal{L}_{tri} = \max(0, ||\mathbf{v}_A - \mathbf{v}_P||_2^2 - ||\mathbf{v}_A - \mathbf{v}_N||_2^2 + \alpha)
\end{equation}

Complementing this, the BCE term regularizes the network to output calibrated probabilities. For a prediction $\hat{y}$ and a ground truth label $y$, it is defined as:
\begin{equation}
\mathcal{L}_{BCE}(y, \hat{y}) = - [y \log(\hat{y}) + (1 - y) \log(1 - \hat{y})]
\end{equation}
In Eq. \ref{total_loss}, for the term $\mathcal{L}_{BCE}(A, P)$, we assign $y=1$ (valid pair); for $\mathcal{L}_{BCE}(A, N)$, we assign $y=0$ (irrelevant pair). This hybrid objective ensures that the model learns both the geometric manifold and accurate classification boundaries.

Although the hybrid approach was hypothesized to enforce stricter geometric clustering, our empirical experiments demonstrated that the model achieves optimal performance when $\lambda_1=0$ and $\lambda_2=1.0$. This suggests that for our specific task, the probabilistic calibration provided by BCE Loss is sufficient and more effective than the distance-based constraints. Consequently, the final model is trained using the BCE objective, ensuring accurate probabilistic alignment between the hyperlink and the target webpage.

\subsection{Weighted Similarity Aggregation}
A hyperlink $H$ contains multiple text features ($H_{main}$, $H_{side1}$, etc.), and a webpage $P$ contains multiple features ($P_{title}$, $P_{header}$, etc.). Simply comparing anchor text to the page title is insufficient.
SemLink computes the similarity score for every pair-wise combination $(T_{H_i}, T_{P_j})$ between the source features and target features. The final relevance score $S_{final}$ is derived using a position-aware weighting scheme:
\begin{equation}
S_{final} = \max_{i, j} ( \text{Model}(T_{H_i}, T_{P_j}) \times w(T) )
\end{equation}
We employ the maximization operator ($\max$) rather than averaging. This alignment follows the Multiple Instance Learning paradigm ~\cite{maron1998framework}, where a bag of instances is considered positive if at least one constituent instance is positive. In the context of web verification, a hyperlink's semantic intent is often localized to a single explicit cue (e.g., the anchor text), while surrounding nodes may be generic noise. Maximization ensures that such a single strong signal is sufficient for validation, preventing the relevant cue from being diluted by low-relevance background text as would occur with averaging ~\cite{collobert2011natural}.
The term $w(T)$ is a spatial decay factor based on the DOM distance from the anchor. We assign maximal weight $w(T)=1.0$ for Anchor Text and Image OCR, and linear decay weights ($0.9, 0.8, \dots, 0.5$) for Side-Texts 1 through 5. This heuristic is grounded in the Gestalt Law of Proximity ~\cite{wertheimer1923laws}, which posits that spatially adjacent elements form a cohesive semantic unit, a principle further supported by Vision-based Page Segmentation ~\cite{cai2003vips}. Consequently, text structurally closer to the interaction point contributes more to the prediction. Finally, the link is classified as \textbf{Valid} if $S_{final} \ge \tau$ (an empirically determined threshold), and \textbf{Irrelevant} otherwise.


\section{Experimental Setup}
\label{sec:experiments}

To evaluate the efficacy, robustness, and efficiency of SemLink as an automated test oracle, we designed a comprehensive suite of experiments guided by the following Research Questions (RQs):

\begin{itemize}
    \item \textbf{RQ1 (Effectiveness):} How does the performance of SemLink compare to state-of-the-art Large Language Models (LLMs) in identifying semantic link rot? 
    \item \textbf{RQ2 (Ablation):} What are the individual contributions of specific feature extraction components (e.g., Side-Text, Image OCR) and the hybrid loss function to the overall model performance? 
    \item \textbf{RQ3 (Efficiency):} Is SemLink computationally efficient enough to be deployed in high-frequency Continuous Integration (CI) environments, specifically in comparison to LLM-based approaches?
\end{itemize}

\subsection{Datasets and Splits}
\label{subsec:dataset}
We utilized the \textbf{HWPPs Dataset} detailed in Section \ref{sec:dataset}. The dataset was stratified by language (Chinese/English) and randomly partitioned into a Training Set (55,000 pairs) for model fine-tuning and a Validation Set (8,870 pairs) for hyperparameter optimization.
To evaluate the model's generalization capability on unseen data, we collected an additional Independent Test Set from 100 real-world webpages distinct from the training domains. These pages yielded 16,951 raw hyperlinks. From this pool, we constructed a balanced evaluation dataset of 4,000 pairs (comprising 2,000 positive and 2,000 negative samples) through manual annotation. This independent set serves as the ground truth for our comparative analysis in RQ1.

\subsection{LLM Baselines}

To benchmark semantic understanding against the state-of-the-art, we selected five representative Large Language Models (LLMs): \textbf{GPT-5.2} (OpenAI's latest flagship with advanced reasoning capabilities) and \textbf{GPT-4o} (the previous commercial SOTA) represent the upper bound of generative performance. We also include \textbf{GPT-3.5 Turbo} as a cost-effective commercial baseline, alongside the open-source \textbf{Llama-3} family (8B and 70B variants) to evaluate performance under resource-constrained deployment scenarios.

To query these LLMs, we employed a structured zero-shot prompt template designed to elicit a scalar relevance score. The prompt structure is as follows:

\footnotesize
\begin{quote}
\textit{``Assume you are a webpage visitor. You expect to see relevant webpage content when clicking on a hyperlink.\\
---\\
Your task is to determine the answer to the following question based on the rating criteria provided.\\
Q: After a webpage visitor clicks on a hyperlink with "Hyperlink Information," do they expect to view a webpage with "Target Webpage Information"?\\
--- Rating Criteria ---\\
1 - Definitely not\\
2 - Probably not\\
3 - Might or might not\\
4 - Probably yes\\
5 - Definitely yes\\
---\\
"Hyperlink Information":\\
\{link\_ info\}\\
---\\
"Target Webpage Information":\\
\{webpage\_info\}\\
---\\
Just give me a rating and do not reply with anything else. Your reply should only be in the following format:\\
``Rating criteria: \textless rating criteria\textgreater''''}
\end{quote}

\normalsize
The models were instructed to output a rating based on a 5-point Likert scale, ranging from \textit{1 - Definitely not} to \textit{5 - Definitely yes}. We explicitly constrained the output format to facilitate automated parsing.

\subsection{Implementation Details}
SemLink was implemented using Python 3.12 and PyTorch 2.2.
For hardware, all local models (SemLink, Llama-3) were evaluated on a workstation equipped with an Intel Core i7-14700K CPU and a single NVIDIA GeForce RTX 4090 GPU (24GB VRAM).
For training, we ran the model for 200 epochs using the Adam optimizer with an initial learning rate of $1e^{-3}$, applying a decay factor of 0.8 every 50 epochs. Hyperparameter optimization was performed via grid search on the validation set (see Appendix \ref{app:hyperparameters} for sensitivity analysis).
For the loss function, the total loss combines Triplet Margin Loss and BCE. The weights were empirically set to $\lambda_1=0$ and $\lambda_2=1.0$, with a triplet margin of $\alpha=1.0$.

\section{Results and Evaluation}
\label{sec:results}

In this section, we present the experimental results answering our three research questions. We analyze the effectiveness of SemLink as a semantic test oracle, determining the impact of our specific feature extraction strategies and quantifying the efficiency gains over generative AI approaches.

\subsection{RQ1: Effectiveness vs. Baselines}
To answer RQ1, we evaluated SemLink and the baseline models on the Independent Test Set described in Section \ref{subsec:dataset} (comprising 4,000 annotated pairs from 100 webpages). Table \ref{tab:rq1_comparison} summarizes the quantitative performance results.

\begin{table}[htbp]
\caption{Performance Comparison with Baselines on Test Set}
\begin{center}
\begin{tabular}{lcccc}
\toprule
\textbf{Model/Tool} & \textbf{Accuracy} & \textbf{Precision} & \textbf{Recall} & \textbf{F1-Score} \\
\multicolumn{5}{l}{\textit{Generative Oracles (LLMs)}} \\
Llama-3-8B & 86.98\% & 80.70\% & 97.20\% & 88.18\% \\
Llama-3-70B & 95.37\% & 95.44\% & 95.30\% & 95.37\% \\
GPT-3.5 Turbo & 94.55\% & 99.37\% & 89.65\% & 94.27\% \\
GPT-4o & 98.40\% & \textbf{99.95\%} & 96.85\% & 98.37\% \\
GPT-5.2 & \textbf{99.10\%} & 99.90\% & \textbf{98.30\%} & \textbf{99.09\%} \\
\midrule
\multicolumn{5}{l}{\textit{Proposed Method}} \\
\textbf{SemLink (Ours)} & 92.70\% & 90.06\% & 96.00\% & 92.93\% \\
\bottomrule
\end{tabular}
\label{tab:rq1_comparison}
\end{center}
\end{table}

Table \ref{tab:rq1_comparison} presents a comprehensive performance comparison between our proposed SemLink and various Generative Oracles, ranging from lightweight open-source models to advanced proprietary systems. SemLink achieves a competitive F1-Score of 92.93\%, establishing a substantial margin over the comparable open-source baseline, Llama-3-8B (88.18\%). While advanced proprietary models such as GPT-4o and the state-of-the-art GPT-5.2 set a high performance ceiling (achieving 98.37\% and 99.09\% respectively), SemLink demonstrates remarkable efficiency by narrowing the performance gap without relying on extreme parameter scales.

Most critically, SemLink exhibits superior sensitivity. As shown in Table \ref{tab:rq1_comparison}, our method achieves a Recall of 96.00\%, explicitly outperforming GPT-3.5 Turbo (89.65\%) and edging out the significantly larger Llama-3-70B (95.30\%). This indicates that while generalized LLMs may dominate in overall precision, SemLink’s specialized architecture is uniquely effective at minimizing false negatives, making it a robust solution for verification tasks where recall is paramount.

\subsection{RQ2: Ablation Study}
To understand which components contribute to SemLink's success, we conducted an ablation study on the feature extraction pipeline and model architecture.

\subsubsection{Impact of Contextual Features}
We trained variants of SemLink using different subsets of $H_{info}$. The results are presented in Table \ref{tab:ablation_features}.

\begin{table}[htbp]
\caption{Ablation Study: Impact of Feature Extraction Strategies}
\begin{center}
\begin{tabular}{lcccc}
\toprule
\textbf{Feature Configuration} & 
\textbf{Accuracy} &
\textbf{Precision} & \textbf{Recall} & \textbf{F1-Score} \\
\midrule
Anchor Text Only & 92.03\% & \textbf{94.45\%} & 89.30\% & 91.80\% \\
Anchor + Side-Text & 92.65\% & 90.31\% & 95.56\% & 92.86\% \\
Anchor + Image OCR & 92.28\% & 94.11\% & 90.20\% & 92.11\% \\
\textbf{Full Context (SemLink)} & \textbf{92.70\%} & 90.06\% & \textbf{96.00\%} & \textbf{92.93\%} \\
\bottomrule
\end{tabular}
\label{tab:ablation_features}
\end{center}
\end{table}

Using only Anchor Text yields a high precision but significantly lower Recall (89.30\%). This confirms our hypothesis that generic anchors (e.g., "Click here") are insufficient for semantic matching. Adding Side-Text improves Recall by +6.26\%, illustrating the value of DOM-based context. The combination of all features (Anchor + Side + Image) yields the best balance, demonstrating that multi-modal extraction is essential for robust web testing.

\subsubsection{Impact of Image Processing}
Focusing specifically on hyperlinks containing images (2,406 samples), we evaluated the contribution of different image text extraction methods (Table \ref{tab:ablation_images}). The combination of HTML Attributes (\texttt{alt}/\texttt{title}) and OCR yields the highest F1-score (90.60\%). Interestingly, using AI-based icon labeling (LabelDroid) slightly degraded performance, likely due to the noise introduced by mislabeling abstract icons.

\begin{table}[htbp]
\caption{Ablation: Image Text Extraction Methods (Image Links Only)}
\begin{center}
\begin{tabular}{lc}
\toprule
\textbf{Method} & \textbf{F1-Score} \\
\midrule
No Image Text (Baseline) & 84.07\% \\
Attributes Only (\texttt{alt}/\texttt{title}) & 89.86\% \\
OCR Only & 86.46\% \\
\textbf{Attributes + OCR} & \textbf{90.60\%} \\
Attributes + OCR + LabelDroid & 88.81\% \\
\bottomrule
\end{tabular}
\label{tab:ablation_images}
\end{center}
\end{table}

\subsection{RQ3: Efficiency and Scalability}
For a test oracle to be practical in CI/CD pipelines, inference speed and cost are critical. We measured the throughput (links processed per second) on a single NVIDIA RTX 4090 GPU. The results are depicted in Table \ref{tab:efficiency}.

\begin{table}[htbp]
\caption{Efficiency Comparison: Throughput vs. Recall}
\begin{center}
\begin{tabular}{lccr}
\toprule
\textbf{Model} & \textbf{Recall} & \textbf{Speed (links/sec)} & \textbf{Speedup} \\
\midrule
Llama-3-8B & 97.20\% & 0.87 & 8.7x \\
Llama-3-70B & 95.30\% & 0.10 & 1x \\
GPT-3.5 (API) & 89.65\% & 1.27 & 12.7x \\
GPT-4o (API) & 96.85\% & 1.17 & 11.7x \\
GPT-5.2 (API) & 98.30\% & 0.65 & 6.5x \\
\textbf{SemLink} & \textbf{96.00\%} & \textbf{30.87} & \textbf{308x} \\
\bottomrule
\end{tabular}
\label{tab:efficiency}
\end{center}
\end{table}

SemLink processes approximately 30.87 links per second, which is roughly \textbf{47.5 times faster} than GPT-5.2 and \textbf{300 times faster} than a locally hosted Llama-3-70B. To visualize this trade-off, we plot the efficiency landscape in Fig. \ref{fig:GreenAi}.

\subsubsection{The Real-time Threshold ($<0.1$s)}
\begin{figure}
    \centering
    \includegraphics[width=1\linewidth]{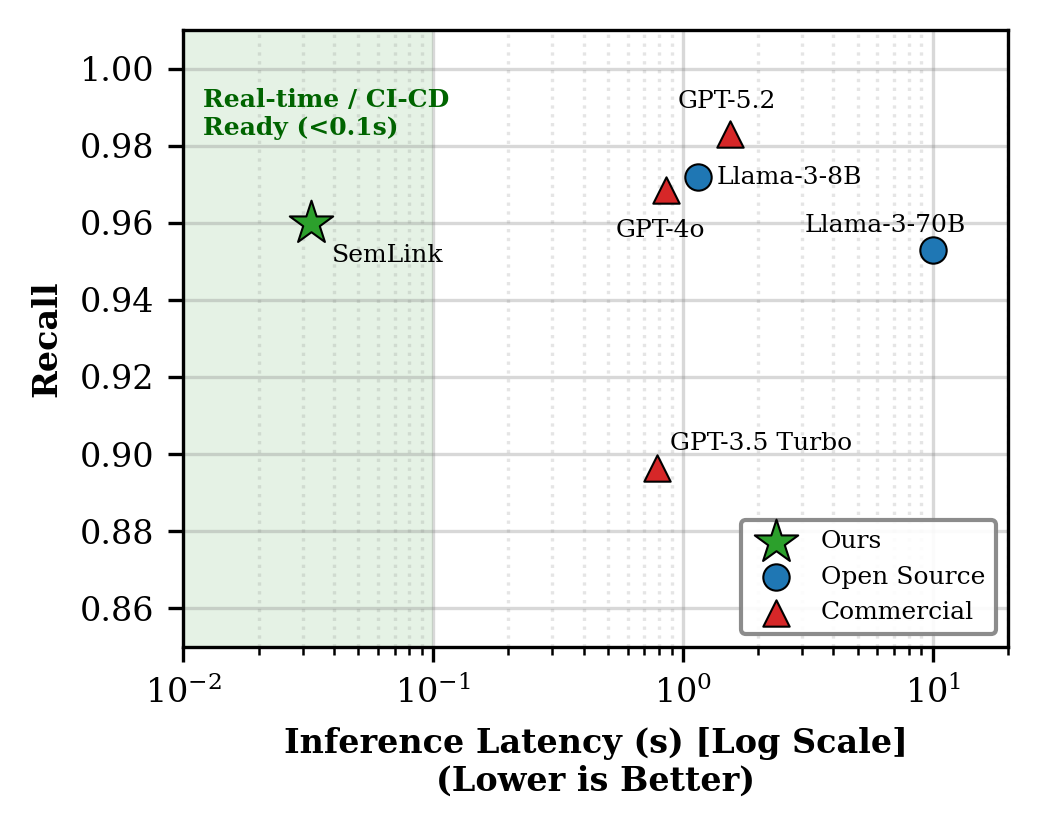} 
    \caption{Efficiency vs. Performance. SemLink (Green Star) is the sole approach within the real-time threshold ($<0.1$s).}
    \label{fig:GreenAi}
\end{figure}

The green shaded region in Fig. \ref{fig:GreenAi} represents the ``Real-time / CI-CD Ready'' zone. The 0.1s threshold is critical for two reasons: First, according to HCI standards, 0.1s is the limit for a system response to be perceived as instantaneous~\cite{nielsen1993usability}. Second, for a regression suite of 100,000 links:
\begin{itemize}
    \item \textbf{SemLink:} $\approx$ \textbf{54 minutes} (Local, magnitude lower cost).
    \item \textbf{GPT-5.2:} $\approx$ \textbf{42.7 hours} (API latency) + estimated \$146 USD cost.
\end{itemize}
SemLink allows for daily or even per-commit verification, whereas LLM-based approaches are computationally prohibitive for frequent regression testing.

\subsubsection{End-to-End Latency Analysis}
It is important to note that the above figures strictly represent model inference throughput. In a real-world end-to-end deployment, the actual bottleneck is typically dominated by the network latency required for web crawling. However, this observation actually \textbf{strengthens} the case for SemLink. By reducing the inference time to an almost negligible fraction ($\approx 0.03$s), SemLink ensures that the testing process remains bound only by unavoidable network constraints. Conversely, using LLM-based approaches would add a significant computational overhead on top of the network latency, effectively doubling the wait time for developers.

\subsection{Qualitative Case Studies}
\label{sec:qualitative}

To demonstrate SemLink's capability as a functional test oracle compared to traditional crash oracles, we analyze specific real-world scenarios encountered during our testing of 30 live websites.

\subsubsection{Case 1: The ``Soft 404'' Failure}
\begin{figure}[!htbp]
    \centering
    \includegraphics[width=0.95\linewidth]{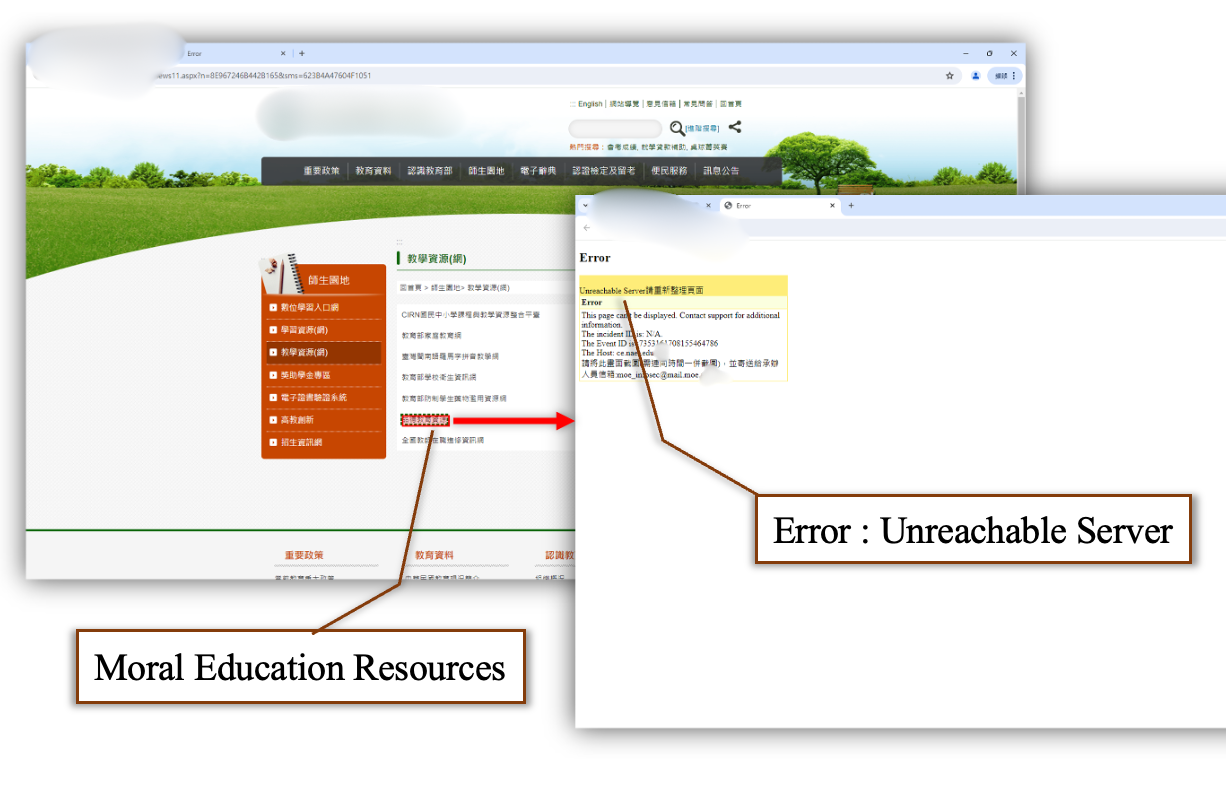}
    \caption{A ``Soft Link Rot'' example. The server returns HTTP 200, but the content is an error message. Traditional tools mark this as Valid; SemLink correctly marks it as Irrelevant.}
    \label{fig:soft_404}
\end{figure}

Traditional tools rely almost exclusively on HTTP status codes. Fig.~\ref{fig:soft_404} illustrates a scenario encountered on a government education portal. The hyperlink pointing to ``Moral Education Resources'' returned a valid \texttt{HTTP 200 OK} status. However, the rendered page displayed a database application error message (``Unreachable Server'').

\begin{itemize}
    \item \textbf{Traditional Tool Result:} Valid (Pass).
    \item \textbf{SemLink Result:} Irrelevant (Fail).
\end{itemize}
SemLink correctly identified that the semantic content of an error message had no coherence with the anchor text, successfully acting as a functional oracle where traditional tools failed.

\subsubsection{Case 2: content mismatch}
\begin{figure}[!htbp]
    \centering
    \includegraphics[width=0.95\linewidth]{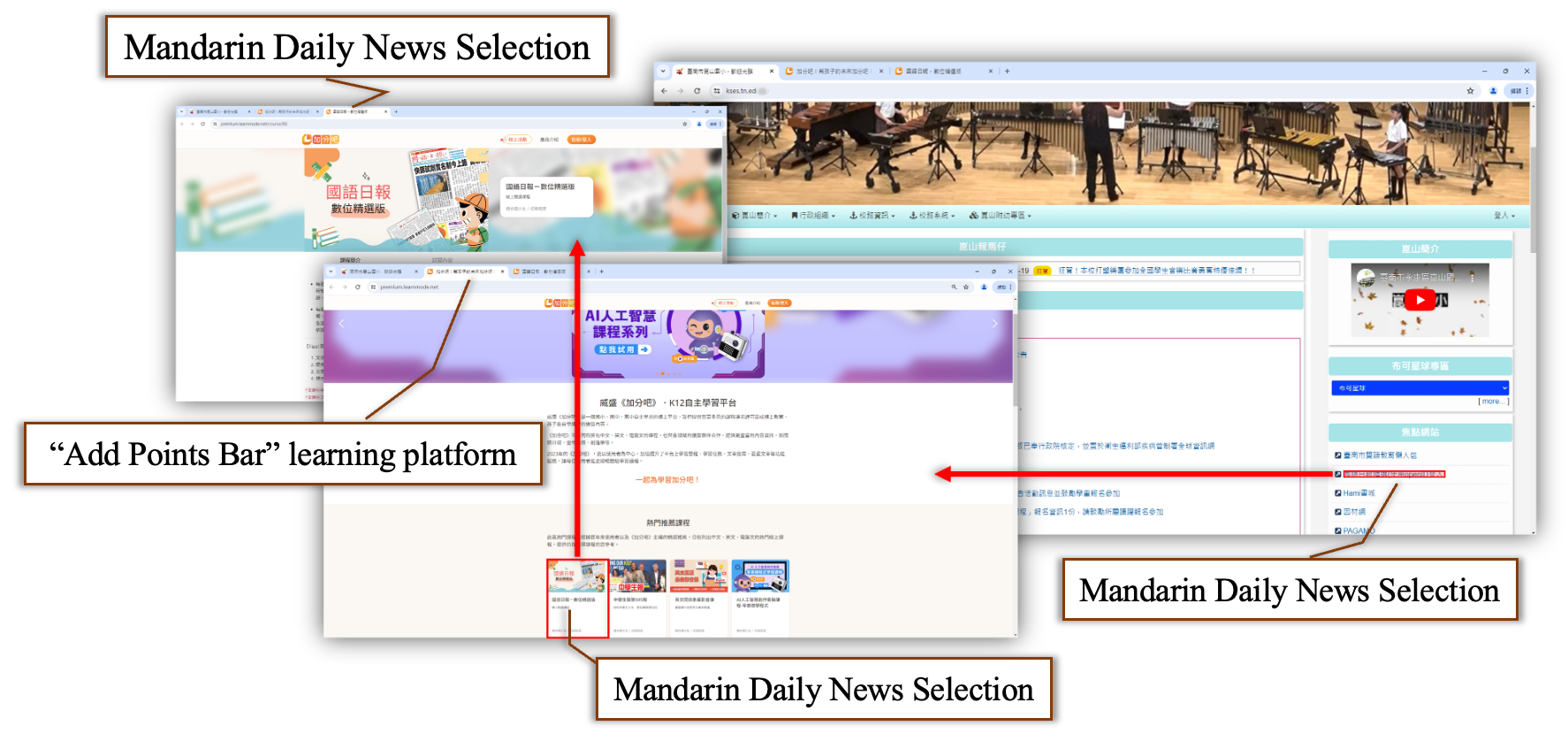}
    \caption{An example of Semantic Drift. The anchor text promises specific news content, but the target page has drifted to a general landing page. SemLink identifies this as Irrelevant due to low semantic overlap.}
    \label{fig:content_mismatch}
\end{figure}

Fig. \ref{fig:content_mismatch} demonstrates a case of content mismatch. The anchor text promised ``Mandarin Daily News Selection,'' but the target URL redirected to a generalized learning platform (``Add Points Bar''). Although the target page included a small sidebar mentioning the newspaper, the overall content was largely irrelevant.

This highlights SemLink's sensitivity to the \textit{primary} topic of the page, leading users to content that is only tangentially related.

\subsubsection{Case 3: Context Disambiguation}
\begin{figure}[!htbp]
    \centering
    \includegraphics[width=0.95\linewidth]{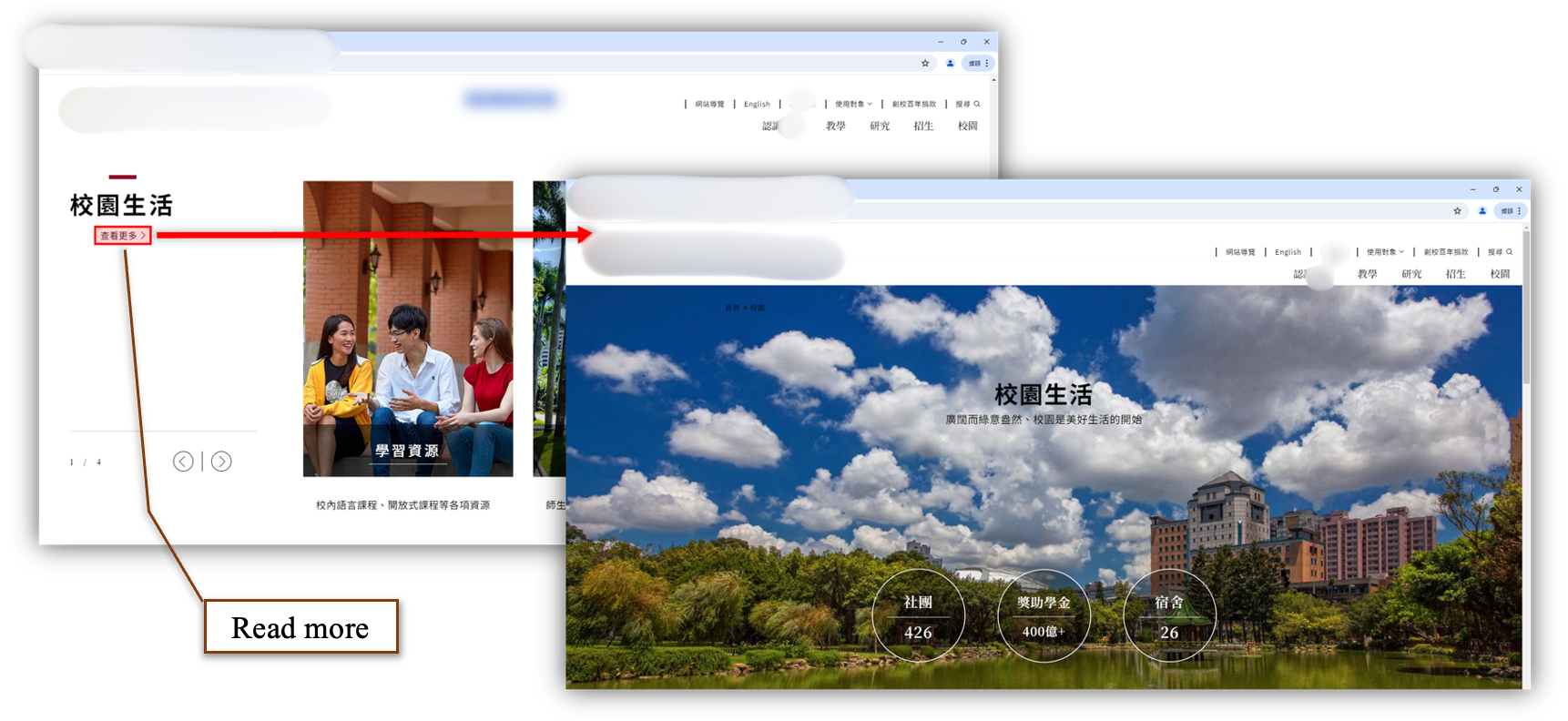}
    \caption{Visualizing the Side-Text Heuristic. SemLink extracts the parent header ``Campus Life'' to disambiguate the generic ``Read More'' anchor text, correctly validating the link.}
    \label{fig:context_disambiguation}
\end{figure}

A common challenge in web testing is generic anchor text. As shown in Fig. \ref{fig:context_disambiguation}, a link labeled simply ``Read More'' was correctly validated by SemLink.

\begin{itemize}
    \item \textbf{Mechanism:} By extracting the Side-Text ``Campus Life'' from the parent DOM element (weight=0.9), SemLink constructed a context vector representing ``Campus Life Read More.''
    \item \textbf{Result:} Valid (Pass).
\end{itemize}
Without the heuristic weighting of Side-Text, this link would have been a False Negative.

\subsection{Error Analysis}
\label{sec:errors}

To understand the limitations of our approach, we analyzed the False Negatives (valid links marked as Irrelevant) encountered during testing. We identified three primary failure modes.

\subsubsection{Login and Authorization Redirects}
\begin{figure}[!htbp]
    \centering
    \includegraphics[width=0.95\linewidth]{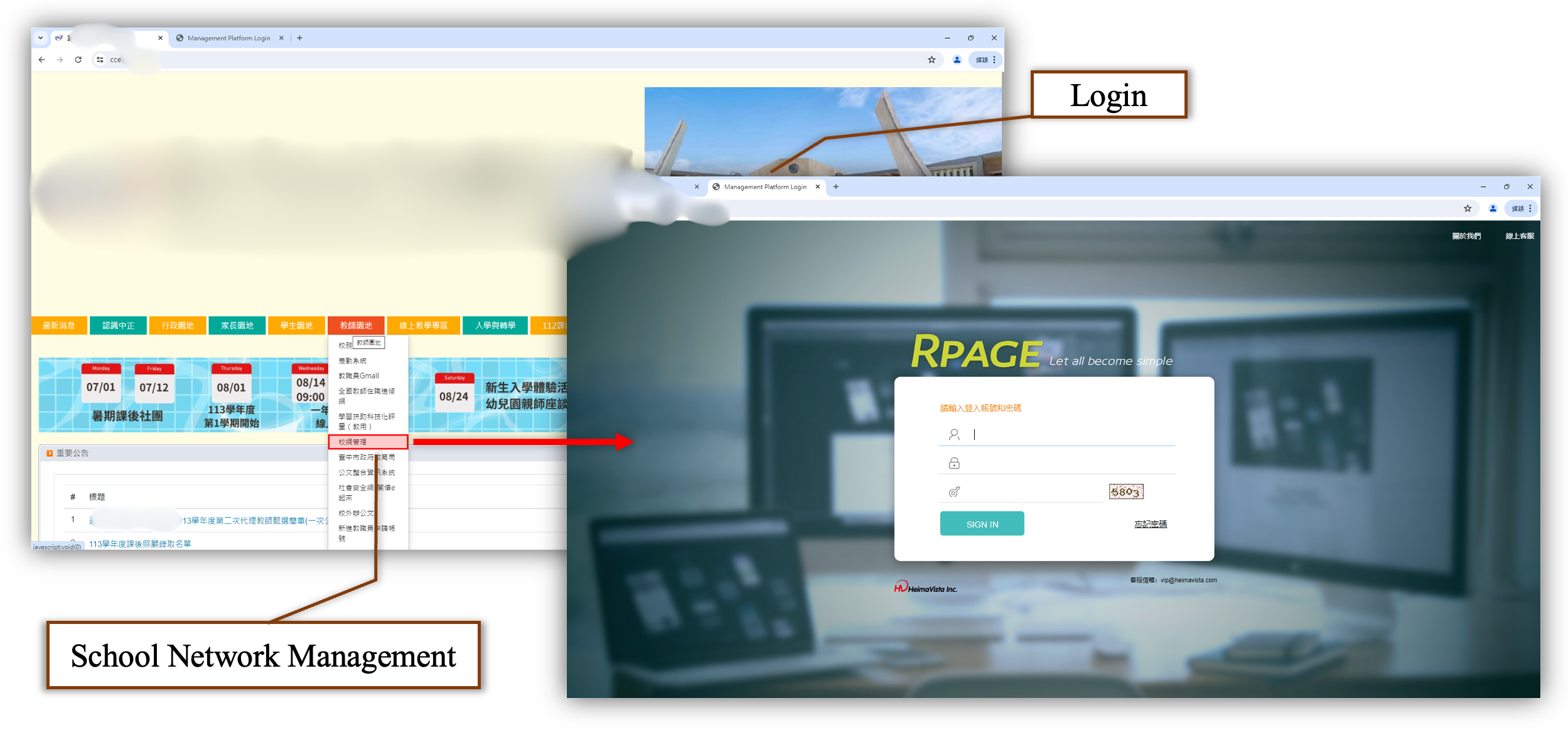}
    \caption{A False Negative case. The link redirects to a login portal. SemLink classifies this as ``Irrelevant'' due to lack of semantic overlap, though it is functional behavior.}
    \label{fig:login_error}
\end{figure}

The most common cause of False Negatives was unexpected redirects to authentication portals. As shown in Fig. \ref{fig:login_error}, a link labeled ``School Network Management'' redirected to a generic system login page.

SemLink correctly determined that the semantic content of a login form does not match the source text ``Management.'' While technically a semantic mismatch, this is often intended behavior. Future work could incorporate a ``Login Page Classifier'' to handle these functional redirects.

\subsubsection{Visual-Only Targets}
SemLink relies on textual features. We encountered failures where the target page conveyed information almost exclusively through images without accessible text or \texttt{alt} attributes (e.g., a library homepage displaying book covers as banners). In these cases, the extracted $P_{info}$ was sparse, leading to low similarity scores. This confirms the need for integrating Vision-Language Models (VLMs) for multimodal verification.

\subsubsection{Generic Category Labels}
Links serving as high-level category filters (e.g., ``Announcements'') often point to lists of specific items (e.g., ``Typhoon Day Off,'' ``New Semester Schedule''). SemLink sometimes marked these as irrelevant because the specific child items did not semantically match the broad parent category string.

\section{Discussion}
\label{sec:discussion}

Our results suggest that specialized discriminative models like SemLink can serve as effective surrogates for expensive generative oracles in high-frequency testing environments.

\subsection{Implications for Web Engineering}
The distinction between "Crash Oracles" (HTTP 404) and "Semantic Oracles" (SemLink) is critical for modern web engineering.
\begin{itemize}
    \item \textbf{Automated Regression Testing:} SemLink processes $\approx$30 links/sec, enabling it to run within standard nightly CI/CD pipelines for large enterprise websites (10k+ pages). This capability allows teams to detect "soft rot" immediately after deployment, preventing semantic drift from accumulating over years.
    \item \textbf{Legacy Migration:} When migrating content management systems (CMS), URLs often change structure. SemLink can automatically verify if the new redirects point to semantically equivalent content, a task currently performed manually or via fragile regex rules.
\end{itemize}

\subsection{Threats to Validity}
We acknowledge several threats to the validity of our study. Regarding internal validity, the heuristic weights for Side-Text extraction were determined empirically and may not be optimal for all web layouts. However, our analysis indicates that the semantic influence of DOM context is highly localized. Since valid classifications require a score above the threshold $\tau=0.7$, contextual nodes at distances $d \ge 4$ (where weights $w < 0.7$) are mathematically excluded from independently triggering a positive decision. Consequently, while we currently consider $k=5$ neighbors for robustness, the effective receptive field is limited to the immediate three nodes. This suggests that the empirical weights successfully filter out distant noise, and future work could explicitly prune the extraction window to $k=3$ for efficiency.
Concerning external validity, although the HWPPs dataset \textbf{provides a substantial corpus of over 60,000 pairs}, it primarily comprises structured domains (e.g., news, government portals). Consequently, SemLink's generalization to highly unstructured or WebGL-based interfaces remains unverified. 
Finally, with respect to construct validity, "semantic relevance" is inherently subjective. We mitigated this by employing human annotation on the independent Test Set to establish ground truth, though individual user expectations for link destinations may still vary.

\section{Conclusion}
\label{sec:conclusion}

This paper presented \textbf{SemLink}, an automated test oracle designed to solve the problem of semantic drift in webpage hyperlinks. Unlike traditional tools that only check for connection failures, SemLink utilizes a Siamese Sentence-BERT architecture to verify that the \textit{content} of a target page fulfills the \textit{semantic promise} of its source anchor.

We introduced the \textbf{HWPPs Dataset}, a rigorous corpus of over 60,000 positive pairs, to facilitate supervised training of semantic navigation models. Our experimental evaluation demonstrated that SemLink achieves a Recall of 96.00\%, effectively matching the performance of commercial Large Language Models like GPT-5.2 while operating at a fraction of the cost and latency.

Future work will focus on integrating multimodal LLMs to better understand complex visual layouts and developing graph neural networks to model site-wide semantic flow rather than individual link pairs. By bridging the gap between syntactic checking and semantic understanding, SemLink represents a significant step towards fully autonomous web quality assurance.

\appendices

\section{Hyperparameter Sensitivity Analysis}
\label{app:hyperparameters}

To ensure the robustness of SemLink, we performed a grid search to optimize the hyperparameters of the Siamese Network and the Hybrid Loss function. The search space and selected optimal values are detailed in Table \ref{tab:hyperparameters}.
\begin{table}[htbp]
\caption{Hyperparameter Search Space and Optimal Values}
\centering
\scriptsize
\renewcommand{\arraystretch}{0.65}
\begin{tabular}{lll}
\toprule
\textbf{Parameter} & \textbf{Search Space} & \textbf{Optimal} \\
\midrule
Triplet Margin ($\alpha$) & $\{0.5, 1.0, 2.0\}$ & $1.0$ \\
Triplet Weight ($\lambda_1$) & $\{0, 0.5, 1.0\}$ & $0$ \\
BCE Weight ($\lambda_2$) & $\{0.5, 1.0, 2.0\}$ & $1.0$ \\
Dense Layers & $\{1, 2, 3\}$ & $2$ \\
\bottomrule
\end{tabular}
\label{tab:hyperparameters}
\end{table}

We observed that the model is particularly sensitive to the Triplet Margin $\alpha$. A small margin ($\alpha < 0.5$) failed to push irrelevant pairs far enough apart, resulting in high False Positives. Conversely, a large margin ($\alpha > 2.0$) made convergence difficult during training. Furthermore, we found that the model performs best when the Triplet Loss is deactivated (i.e., $\lambda_1 = 0$ and $\lambda_2 = 1$). We hypothesize that the efficacy of the Triplet Loss is constrained by the limited batch size, which reduces the diversity of negative samples, or by vanishing gradients absent effective hard negative mining. Consequently, the BCE Loss proves to be more robust, providing a stable gradient flow for feature discrimination.

\section{Heuristic Weighting Function}
\label{app:weighting}

As described in Section \ref{sec:model_architecture}, SemLink utilizes a heuristic weighting function to aggregate similarity scores from different context sources. The weights were determined based on a preliminary analysis of the DOM distance relative to the anchor tag.

Let $d(T)$ be the node distance of a text element $T$ from the anchor tag $A$ in the DOM tree. We define the weight $w(T)$ as:
\begin{equation}
\footnotesize
w(T) = \begin{cases} 
1.0 & \text{if } T \in \{T_{anchor}, T_{img\_ocr}\} \\
1.0 - 0.1 \cdot k & \text{if } T = \text{Side-Text}_k, \, k \in \{1, \dots, 5\} \\
0 & \text{otherwise}
\end{cases}
\end{equation}

The decay reflects the intuition that textual relevance decreases as we move further away from the interaction point. For example, ``Side-Text 1" usually corresponds to a button label or a direct card description, whereas ``Side-Text 5" might be a generic section header or unrelated footer text.

\bibliographystyle{IEEEtran}
\bibliography{references}

@inproceedings{reimers2019sentence,
  author={Reimers, Nils and Gurevych, Iryna},
  title        = {Sentence-BERT: Sentence Embeddings using Siamese BERT-Networks},
  booktitle    = {Proceedings of the 2019 Conference on Empirical Methods in Natural
                  Language Processing and the 9th International Joint Conference on
                  Natural Language Processing, {EMNLP-IJCNLP} 2019, Hong Kong, China,
                  November 3-7, 2019},
  pages        = {3980--3990},
  publisher    = {Association for Computational Linguistics},
  year         = {2019},
  url          = {https://doi.org/10.18653/v1/D19-1410},
  doi          = {10.18653/V1/D19-1410}
}

@inproceedings{devlin2018bert,
  author       = {Jacob Devlin and
                  Ming{-}Wei Chang and
                  Kenton Lee and
                  Kristina Toutanova},
  editor       = {Jill Burstein and
                  Christy Doran and
                  Thamar Solorio},
  title        = {{BERT:} Pre-training of Deep Bidirectional Transformers for Language Understanding},
  booktitle    = {Proceedings of the 2019 Conference of the North American Chapter of the Association for Computational Linguistics: Human Language Technologies,
                  {NAACL-HLT} 2019, Minneapolis, MN, USA, June 2-7, 2019, Volume 1 (Long
                  and Short Papers)},
  pages        = {4171--4186},
  publisher    = {Association for Computational Linguistics},
  year         = {2019},
  url          = {https://doi.org/10.18653/v1/n19-1423},
  doi          = {10.18653/V1/N19-1423}
}

@inproceedings{chen2020unblind,
  author={Chen, Jieshan and Chen, Chunyang and Xing, Zhenchang and Xu, Xiwei and Zhu, Liming and Li, Guoqiang and Wang, Jin},
  title        = {Unblind your apps: predicting natural-language labels for mobile {GUI} components by deep learning},
  booktitle    = {{ICSE} '20: 42nd International Conference on Software Engineering, Seoul, South Korea, 27 June - 19 July, 2020},
  pages        = {322--334},
  publisher    = {{ACM}},
  year         = {2020},
  url          = {https://doi.org/10.1145/3377811.3380327},
  doi          = {10.1145/3377811.3380327}
}

@article{han2021survey,
  title={A survey on the techniques, applications, and performance of short text semantic similarity},
  author={Han, Meng and Zhang, Xi and Yuan, Xiaohong and Jiang, Jianhua and Yun, Wen and Gao, Changhua},
  journal={Concurrency and Computation: Practice and Experience},
  volume={33},
  number={5},
  pages={e5971},
  year={2021},
  url          = {https://doi.org/10.1002/cpe.5971},
  doi          = {10.1002/CPE.5971},
}

@article{chicco2021siamese,
  author       = {Davide Chicco},
  editor       = {Hugh M. Cartwright},
  title        = {Siamese Neural Networks: An Overview},
  booktitle    = {Artificial Neural Networks - Third Edition},
  series       = {Methods in Molecular Biology},
  volume       = {2190},
  pages        = {73--94},
  publisher    = {Springer},
  year         = {2021},
  url          = {https://doi.org/10.1007/978-1-0716-0826-5\_3},
  doi          = {10.1007/978-1-0716-0826-5\_3},
}

@article{park2003hyperlink,
  title={{Hyperlink analyses of the World Wide Web: A review}},
  author={Park, Han Woo and Thelwall, Mike},
  journal={Journal of Computer-Mediated Communication},
  volume={8},
  number={4},
  pages={JCMC843},
  year={2003}
}

@article{henzinger2001hyperlink,
  author       = {Monika Rauch Henzinger},
  title        = {Hyperlink Analysis for the Web},
  journal      = {{IEEE} Internet Comput.},
  volume       = {5},
  number       = {1},
  pages        = {45--50},
  year         = {2001},
  url          = {https://doi.org/10.1109/4236.895141},
  doi          = {10.1109/4236.895141}
}

@article{martinez2012updating,
  author       = {Juan Mart{\'{\i}}nez{-}Romo and
                  Lourdes Araujo},
  title        = {Updating broken web links: An automatic recommendation system},
  journal      = {Inf. Process. Manag.},
  volume       = {48},
  number       = {2},
  pages        = {183--203},
  year         = {2012},
  url          = {https://doi.org/10.1016/j.ipm.2011.03.006},
  doi          = {10.1016/J.IPM.2011.03.006}
}

@inproceedings{martinez2008recommendation,
  author       = {Juan Martinez{-}Romo and
                  Lourdes Araujo},
  title        = {Recommendation System for Automatic Recovery of Broken Web Links},
  booktitle    = {Advances in Artificial Intelligence - {IBERAMIA} 2008, 11th Ibero-American
                  Conference on AI, Lisbon, Portugal, October 14-17, 2008. Proceedings},
  series       = {Lecture Notes in Computer Science},
  volume       = {5290},
  pages        = {302--311},
  publisher    = {Springer},
  year         = {2008},
  url          = {https://doi.org/10.1007/978-3-540-88309-8\_31},
  doi          = {10.1007/978-3-540-88309-8\_31},
}

@article{blustein1997methods,
  title={Methods for evaluating the quality of hypertext links},
  author={Blustein, James and Webber, R and Tague-Sutcliffe, Jean},
  journal={Information Processing \& Management},
  volume       = {33},
  number       = {2},
  pages        = {255--271},
  year         = {1997},
  url          = {https://doi.org/10.1016/S0306-4573(96)00066-0},
  doi          = {10.1016/S0306-4573(96)00066-0},
}

@inproceedings{schroff2015facenet,
  author={Schroff, Florian and Kalenichenko, Dmitry and Philbin, James},
  title        = {FaceNet: {A} unified embedding for face recognition and clustering},
  booktitle    = {{IEEE} Conference on Computer Vision and Pattern Recognition, {CVPR}
                  2015, Boston, MA, USA, June 7-12, 2015},
  pages        = {815--823},
  publisher    = {{IEEE} Computer Society},
  year         = {2015},
  url          = {https://doi.org/10.1109/CVPR.2015.7298682},
  doi          = {10.1109/CVPR.2015.7298682}
}

@techreport{chapekis2024online,
  title={When online content disappears},
  author={Chapekis, A and Bestvater, S and Remy, E and Rivero, G},
  institution={Pew Research Center},
  year={2024},
  url={https://www.pewresearch.org/wp-content/uploads/sites/20/2024/05/pl_2024.05.17_link-rot_report.pdf}
}

@misc{deadlinkchecker,
  title={{Dead Link Checker}},
  howpublished={\url{https://www.deadlinkchecker.com/}},
  year={2024},
  note={Accessed: 2025-07-01}
}

@misc{screamingfrog,
  title={{Screaming Frog SEO Spider}},
  howpublished={\url{https://www.screamingfrog.co.uk/seo-spider/}},
  year={2024},
  note={Accessed: 2025-07-01}
}

@misc{whatwg_dom,
  title={{DOM S}tandard},
  author={WHATWG},
  howpublished={\url{https://dom.spec.whatwg.org/}},
  year={2024},
  note={Accessed: 2025-12-10}
}

@article{liu2025relevance,
author = {Liu, Qi and Duan, Haozhe and Mao, Jiaxin and Wen, Ji-Rong},
title = {How do Large Language Models Understand Relevance? A Mechanistic Interpretability Perspective},
year = {2025},
publisher = {Association for Computing Machinery},
address = {New York, NY, USA},
issn = {1046-8188},
url = {https://doi.org/10.1145/3774942},
doi = {10.1145/3774942},
journal = {ACM Trans. Inf. Syst.},
month = nov
}

@inproceedings{Pradel2026Testora,
  author    = {Michael Pradel},
  title     = {Testora: Using Natural Language Intent to Detect Behavioral Regressions},
  booktitle = {2026 IEEE/ACM 48th International Conference on Software Engineering (ICSE '26)},
  year      = {2026},
  address   = {Rio de Janeiro, Brazil},
  publisher = {ACM},
  pages     = {13 pages},
  doi       = {10.1145/3744916.3764527},
  url       = {https://doi.org/10.1145/3744916.3764527}
}

@article{barr2015oracle,
  title={The Oracle Problem in Software Testing: {A} Survey},
  author={Barr, Earl T. and Harman, Mark and McMinn, Phil and Shahbaz, Muzammil and Yoo, Shin},
  journal={IEEE Transactions on Software Engineering},
  volume={41},
  number={5},
  pages={507--525},
  year={2015},
  month={May},
  publisher={IEEE},
  doi={10.1109/TSE.2014.2372785}
}

@article{BALSAM2025112186,
title = {Web application testing—Challenges and opportunities},
journal = {Journal of Systems and Software},
volume = {219},
pages = {112186},
year = {2025},
issn = {0164-1212},
doi = {https://doi.org/10.1016/j.jss.2024.112186},
url = {https://www.sciencedirect.com/science/article/pii/S0164121224002309},
author = {Sebastian Balsam and Deepti Mishra},
}

@inproceedings{qi2023semantic,
  title={Semantic Test Repair for Web Applications},
  author={Qi, Xiaofang and Qian, Xiang and Li, Yanhui},
  booktitle={Proceedings of the 31st ACM Joint European Software Engineering Conference and Symposium on the Foundations of Software Engineering (ESEC/FSE '23)},
  pages={1190--1202},
  year={2023},
  publisher={ACM},
  doi={10.1145/3611643.3616324}
}

@article{Lee24Automated,
author = {Lee, Jaeseong and Chen, Simin and Mordahl, Austin and Liu, Cong and Yang, Wei and Wei, Shiyi},
title = {Automated Testing Linguistic Capabilities of {NLP} Models},
year = {2024},
issue_date = {September 2024},
publisher = {Association for Computing Machinery},
address = {New York, NY, USA},
volume = {33},
number = {7},
issn = {1049-331X},
url = {https://doi.org/10.1145/3672455},
doi = {10.1145/3672455},
journal = {ACM Trans. Softw. Eng. Methodol.},
month = sep,
articleno = {176},
numpages = {33},
  url          = {https://doi.org/10.1145/3672455},
  doi          = {10.1145/3672455},
}

@INPROCEEDINGS{Lin2017Using,
  author={Lin, Jun-Wei and Wang, Farn and Chu, Paul},
  booktitle={2017 IEEE International Conference on Software Testing, Verification and Validation (ICST)}, 
  title={Using Semantic Similarity in Crawling-Based Web Application Testing}, 
  year={2017},
  volume={},
  number={},
  pages={138-148},
  doi={10.1109/ICST.2017.20}}

@article{nass2023similarity,
author = {Nass, Michel and Al\'{e}groth, Emil and Feldt, Robert and Leotta, Maurizio and Ricca, Filippo},
title = {Similarity-based Web Element Localization for Robust Test Automation},
year = {2023},
issue_date = {May 2023},
publisher = {Association for Computing Machinery},
address = {New York, NY, USA},
volume = {32},
number = {3},
issn = {1049-331X},
url = {https://doi.org/10.1145/3571855},
doi = {10.1145/3571855},
journal = {ACM Trans. Softw. Eng. Methodol.},
month = apr,
articleno = {75},
numpages = {30}
}

@INPROCEEDINGS{Kirinuki2021NLP,
  author={Kirinuki, Hiroyuki and Matsumoto, Shinsuke and Higo, Yoshiki and Kusumoto, Shinji},
  booktitle={2021 IEEE International Conference on Software Maintenance and Evolution (ICSME)}, 
  title={NLP-assisted Web Element Identification Toward Script-free Testing}, 
  year={2021},
  volume={},
  number={},
  pages={639-643},
  doi={10.1109/ICSME52107.2021.00072}}

@INPROCEEDINGS{Wen24Enhancing,
  author={Wen, Zhongzhen and Lu, Yifei and Xu, Tongtong and Pan, Minxue and Zhang, Tian and Li, Xuandong},
  booktitle={2024 IEEE International Conference on Software Maintenance and Evolution (ICSME)}, 
  title={Enhancing Web Test Script Repair Using Integrated UI Structural and Visual Information}, 
  year={2024},
  volume={},
  number={},
  pages={75-86},
  doi={10.1109/ICSME58944.2024.00018}}

@inproceedings{Mahajan17Automated,
author = {Mahajan, Sonal and Alameer, Abdulmajeed and McMinn, Phil and Halfond, William G. J.},
title = {Automated repair of layout cross browser issues using search-based techniques},
year = {2017},
isbn = {9781450350761},
publisher = {Association for Computing Machinery},
address = {New York, NY, USA},
url = {https://doi.org/10.1145/3092703.3092726},
doi = {10.1145/3092703.3092726},
booktitle = {Proceedings of the 26th ACM SIGSOFT International Symposium on Software Testing and Analysis},
pages = {249–260},
numpages = {12},
location = {Santa Barbara, CA, USA},
series = {ISSTA 2017}
}

@inproceedings{Huang22AEON,
author = {Huang, Jen-tse and Zhang, Jianping and Wang, Wenxuan and He, Pinjia and Su, Yuxin and Lyu, Michael R.},
title = {AEON: a method for automatic evaluation of NLP test cases},
year = {2022},
isbn = {9781450393799},
publisher = {Association for Computing Machinery},
address = {New York, NY, USA},
url = {https://doi.org/10.1145/3533767.3534394},
doi = {10.1145/3533767.3534394},
pages = {202–214},
numpages = {13},
location = {Virtual, South Korea},
series = {ISSTA 2022}
}

@book{nielsen1993usability,
  author={Nielsen, Jakob},
  title        = {Usability engineering},
  publisher    = {Academic Press},
  year         = {1993},
  isbn         = {978-0-12-518405-2},
}

@inproceedings{mihalcea2004textrank,
    title = "{T}ext{R}ank: Bringing Order into Text",
    author = "Mihalcea, Rada  and
      Tarau, Paul",
    editor = "Lin, Dekang  and
      Wu, Dekai",
    booktitle = "Proceedings of the 2004 Conference on Empirical Methods in Natural Language Processing",
    month = jul,
    year = "2004",
    address = "Barcelona, Spain",
    publisher = "Association for Computational Linguistics",
    url = "https://aclanthology.org/W04-3252/",
    pages = "404--411"
}

@inproceedings{maron1998framework,
  author={Maron, Oded and Lozano-P{\'e}rez, Tom{\'a}s},
  title        = {A Framework for Multiple-Instance Learning},
  booktitle    = {Advances in Neural Information Processing Systems 10, {[NIPS} Conference,
                  Denver, Colorado, USA, 1997]},
  pages        = {570--576},
  publisher    = {The {MIT} Press},
  year         = {1997},
  url          = {http://papers.nips.cc/paper/1346-a-framework-for-multiple-instance-learning},
}

@article{collobert2011natural,
  author       = {Ronan Collobert and
                  Jason Weston and
                  L{\'{e}}on Bottou and
                  Michael Karlen and
                  Koray Kavukcuoglu and
                  Pavel P. Kuksa},
  title        = {Natural Language Processing (Almost) from Scratch},
  journal      = {J. Mach. Learn. Res.},
  volume       = {12},
  pages        = {2493--2537},
  year         = {2011},
  url          = {https://dl.acm.org/doi/10.5555/1953048.2078186},
  doi          = {10.5555/1953048.2078186},
}

@article{wertheimer1923laws,
  title={Laws of organization in perceptual forms.},
  author={Wertheimer, Max},
  year={1938},
  publisher={Kegan Paul, Trench, Trubner \& Company},
url={https://doi.org/10.1037/11496-005},
doi={10.1037/11496-005}
}

@inproceedings{cai2003vips,
  title={{VIPS}: a vision-based page segmentation algorithm},
  author={Cai, Deng and Yu, Shipeng and Wen, Ji-Rong and Ma, Wei-Ying},
  booktitle={Microsoft technical report (MSR-TR-2003-79)},
  year={2003},
  publisher={Microsoft Research}
}

@ARTICLE{yang2025artperception,
  author={Yang, Guan-Yan and Cheng, Tzu-Yu and Teng, Ya-Wen and Wang, Farn and Yeh, Kuo-Hui},
  journal={Journal of Network and Computer Applications}, 
  title={{ArtPerception}: {ASCII} Art-based Jailbreak on LLMs with Recognition
Pre-test}, 
  year={2025},
  url={https://www.sciencedirect.com/science/article/abs/pii/S108480452500253X}
}

@inproceedings{wei2024jailbroken,
  author={Wei, Alexander and Haghtalab, Nika and Steinhardt, Jacob},
  title        = {Jailbroken: How Does {LLM} Safety Training Fail?},
  booktitle    = {Advances in Neural Information Processing Systems 36: Annual Conference
                  on Neural Information Processing Systems 2023, NeurIPS 2023, New Orleans,
                  LA, USA, December 10 - 16, 2023},
  year         = {2023},
  url          = {http://papers.nips.cc/paper\_files/paper/2023/hash/fd6613131889a4b656206c50a8bd7790-Abstract-Conference.html},
}

@book{bishop2006pattern,
  author       = {Christopher M. Bishop},
  title        = {Pattern recognition and machine learning, 5th Edition},
  series       = {Information science and statistics},
  publisher    = {Springer},
  year         = {2007},
  url          = {https://www.worldcat.org/oclc/71008143},
  isbn         = {9780387310732},
}
\end{document}